\documentclass[journal]{IEEEtran}

\usepackage{ifpdf}
\usepackage{cite}
\usepackage{amsmath}
\usepackage{algorithmic}
\usepackage{graphicx}
\usepackage{array}
\usepackage{float}
\usepackage[caption=false,font=normalsize,labelfont=sf,textfont=sf]{subfig}
\usepackage{fixltx2e}
\usepackage{url}
\usepackage{xcolor}
\begin{document}

\title{On Analog Silicon Photomultipliers in Standard 55-nm BCD Technology for LiDAR Applications}

\author{Jiuxuan Zhao,~\IEEEmembership{Student Member,~IEEE,}
        Tommaso Milanese,~\IEEEmembership{Student Member,~IEEE,}
        Francesco Gramuglia,~\IEEEmembership{Student Member,~IEEE,}
        Pouyan Keshavarzian,~\IEEEmembership{Student Member,~IEEE,}
        Shyue Seng Tan,
        Michelle Tng,
        Louis Lim,
        Vinit Dhulla,
        Elgin Quek,~\IEEEmembership{Member,~IEEE,}
        Myung-Jae Lee,~\IEEEmembership{Member,~IEEE,}
        Edoardo Charbon,~\IEEEmembership{Fellow,~IEEE}
        
\thanks{This project has received funding from the European Union's Horizon 2020 research and innovation programme under the Marie Skłodowska-Curie grant agreement No. 754354. This work was supported by the Swiss National Science Foundation under Grant 200021-169465. The work of Pouyan Keshavarzian was supported by Qrypt Inc. (Jiuxuan Zhao and Tommaso Milanese contributed equally to this work.) (Corresponding authors: Myung-Jae Lee, Edoardo Charbon.)
\par Jiuxuan Zhao, Tommaso Milanese, Francesco Gramuglia, Pouyan Keshavarzian and Edoardo Charbon are with the School of Engineering, Ecole Polytechnique Federale de Lausanne (EPFL), 2002 Neuchatel, Switzerland. (e-mail: jiuxuan.zhao@epfl.ch; tommaso.milanese@epfl.ch; francesco.gramuglia@epfl.ch; pouyan.keshavarzian@epfl.ch; edoardo.charbon@epfl.ch).
\par Shyue Seng Tan, Michelle Tng, Louis Lim, Vinit Dhulla, and Elgin Quek are with the GLOBALFOUNDRIES Singapore Pte. Ltd., Singapore 738406, Singapore (e-mail: jason.tan@globalfoundries.com; jinghuamichelle.tng@globalfoundries.
com; louis.lim@globalfoundries.com; vinit.dhulla@globalfoundries.com; elgin.quek@globalfoundries.com).
\par Myung-Jae Lee is with the Post-Silicon Semiconductor Institute, Korea
Institute of Science and Technology (KIST), Seoul 02792, South Korea (e-mail:
mj.lee@kist.re.kr).}}

\maketitle

\begin{abstract}
We present an analog silicon photomultiplier (SiPM) based on a standard 55 nm Bipolar-CMOS-DMOS (BCD) technology. The SiPM is composed of 16$\times$16 single-photon avalanche diodes (SPADs) and  measures 0.29$\times$0.32 mm$^2$. Each SPAD cell is passively quenched by a monolithically integrated 3.3 V thick oxide transistor. The measured gain is 3.4$\times$ 10$^5$ at 5 V excess bias voltage. The single-photon timing resolution (SPTR) is 185 ps and the multiple-photon timing resolution (MPTR) is 120 ps at 3.3 V excess bias voltage. We integrate the SiPM into a co-axial light detection and ranging (LiDAR) system with a time-correlated single-photon counting (TCSPC) module in FPGA. The depth measurement up to 25 m achieves an accuracy of 2 cm and precision of 2 mm under the room ambient light condition. With co-axial scanning, the intensity and depth images of complex scenes with resolutions of 128$\times$256 and 256$\times$512 are demonstrated. The presented SiPM enables the development of cost-effective LiDAR system-on-chip (SoC) in the advanced technology.
\end{abstract}

\begin{IEEEkeywords}
Single-photon avalanche diode, SPAD, silicon photomultiplier, SiPM, CMOS, BCD, time-correlated single-photon counting, TCSPC, LiDAR, co-axial scanning.
\end{IEEEkeywords}

\section{Introduction}

\IEEEPARstart{S}{ilicon} photomultipliers (SiPMs) are large-area photosensors capable of counting individual photons at high timing resolution. A SiPM is implemented as an array of hundreds or thousands of single-photon avalanche diodes (SPADs), called microcells~\cite{buzhan2003silicon, acerbi2019understanding, eckert2010}, that are individually quenched and whose avalanche currents are summed at a single node. SiPMs have been widely used in biomedical imaging~\cite{bruschini2019single}, high-energy physics~\cite{aalseth2018darkside}, optical communications, and light detection and ranging (LiDAR)~\cite{yoshioka201820}. 
There exist at least two types of SiPMs: analog SiPMs (a-SiPMs)~\cite{villa2021spads}, where the avalanche currents are summed in an analog fashion, and digital SiPMs (d-SiPMs), where the avalanche currents are converted to digital signals and combined using logic trees~\cite{gnecchi2016digital} and each microcell can be masked individually~\cite{conca2020large}. 
In an a-SiPM, the output node generates a current proportional to the number of quasi-simultaneously detected photons~\cite{villa2021spads, sanzaro20180}, while a d-SiPM requires additional logic to extract that information, as it is done in a multi-digital SiPM (md-SiPM) \cite{Mandai14}.

In this work, we focus on a-SiPMs for LiDAR applications~\cite{gnecchi2019long, kondo20205,villa2021spads}. By providing the information of the overall received signal intensity, an a-SiPM enables a larger detection area, if compared to a single SPAD, while effectively reducing dead time and maintaining high timing performance. The a-SiPM output however requires dedicated analog processing and timing circuits to extract the time-of-flight of photons as required by LiDAR. A threshold of multiple photons can be set for pulse detection in order to improve the signal background noise ratio (SBNR). Furthermore, the implementation of SiPM based on commercial complementary-metal-oxide-semiconductor (CMOS) technology~\cite{sanzaro20180, zou2015planar} enables  cost-effective systems-on-chip (SoCs) with both detectors and processing electronics~\cite{yoshioka201820,blumino20}.

\section{Device Design}
\begin{figure}[!t]
\centering
\includegraphics[width=1 \columnwidth]{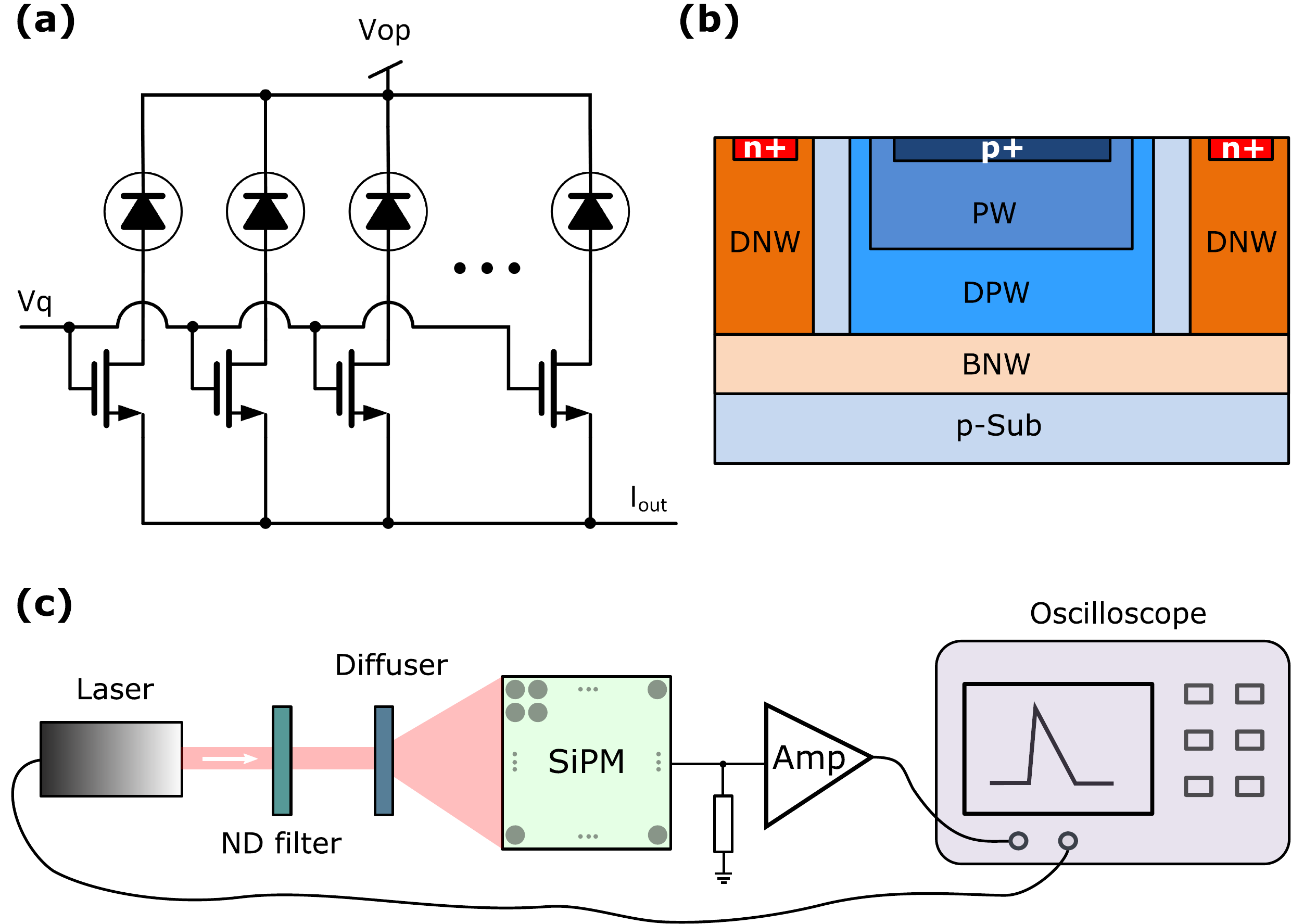}
\caption{(a) Schematic of the SiPM, (b) the cross-section of the SPAD and (c) the measurement setup.}
\label{fig_schematic}
\end{figure}

The SiPM proposed in this work comprises 16$\times$16 SPADs with cathode nodes connected together to bias voltage $V_{op}$, as shown in Fig.~\ref{fig_schematic}a. A 3.3-V thick oxide transistor connected to the anode of each SPAD passively quenches it. The gate nodes of all the transistors are connected together and controlled by an externally applied quenching voltage ($V_q$); they act as quenching resistors. Compared to poly-silicon resistors, the quenching transistors are smaller and their resistance can be adjusted, thus ensuring a variety of operation modes. To achieve better timing jitter, the output current from each cell is connected together into a single pad $I_{out}$ by an \emph{H-tree}.

\begin{figure}
\centering
\includegraphics[width=1 \columnwidth]{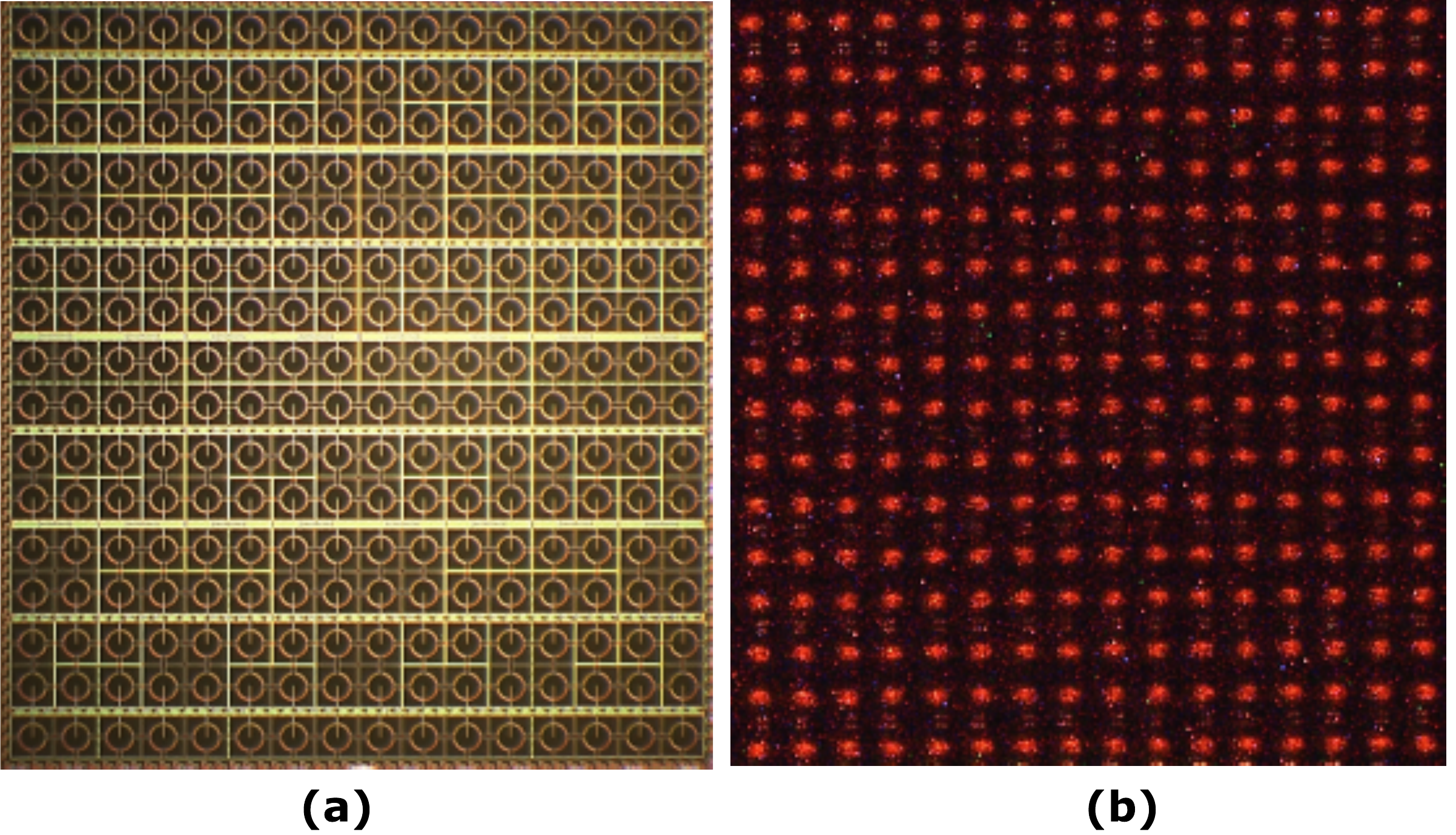}
\caption{Micrograph of the SiPM and light emission test of all the SPAD microcells.}
\label{fig_let}
\end{figure}

As shown in Fig.~\ref{fig_schematic}b, the SPAD cell of the SiPM is based on a \emph{deep p-well / buried n-well} (DPW/BNW) junction. The complete characterization of this SPAD is presented in~\cite{gramuglia}. The entire area of the SiPM is 0.29$\times$0.32 mm$^2$ with 22\% fill factor (FF) and 18.5 $\mu$m pitch. The FF can be improved by mounting a microlens array (MLA). Fig. ~\ref{fig_let}a shows the micrograph of the SiPM array, in which the \emph{H-tree} metal connection can be clearly observed. The light emission test was performed by letting all the SPAD cells firing without quenching ($V_q$ = 3.3 V); the result is displayed in Fig. ~\ref{fig_let}b. The uniform bright region of the array indicates the active area of the SPADs with the absence of premature edge breakdown.\\
\section{Performance Characterization}

\subsection{I-V Characteristics}
\begin{figure}
\centering
\includegraphics[width=1 \columnwidth]{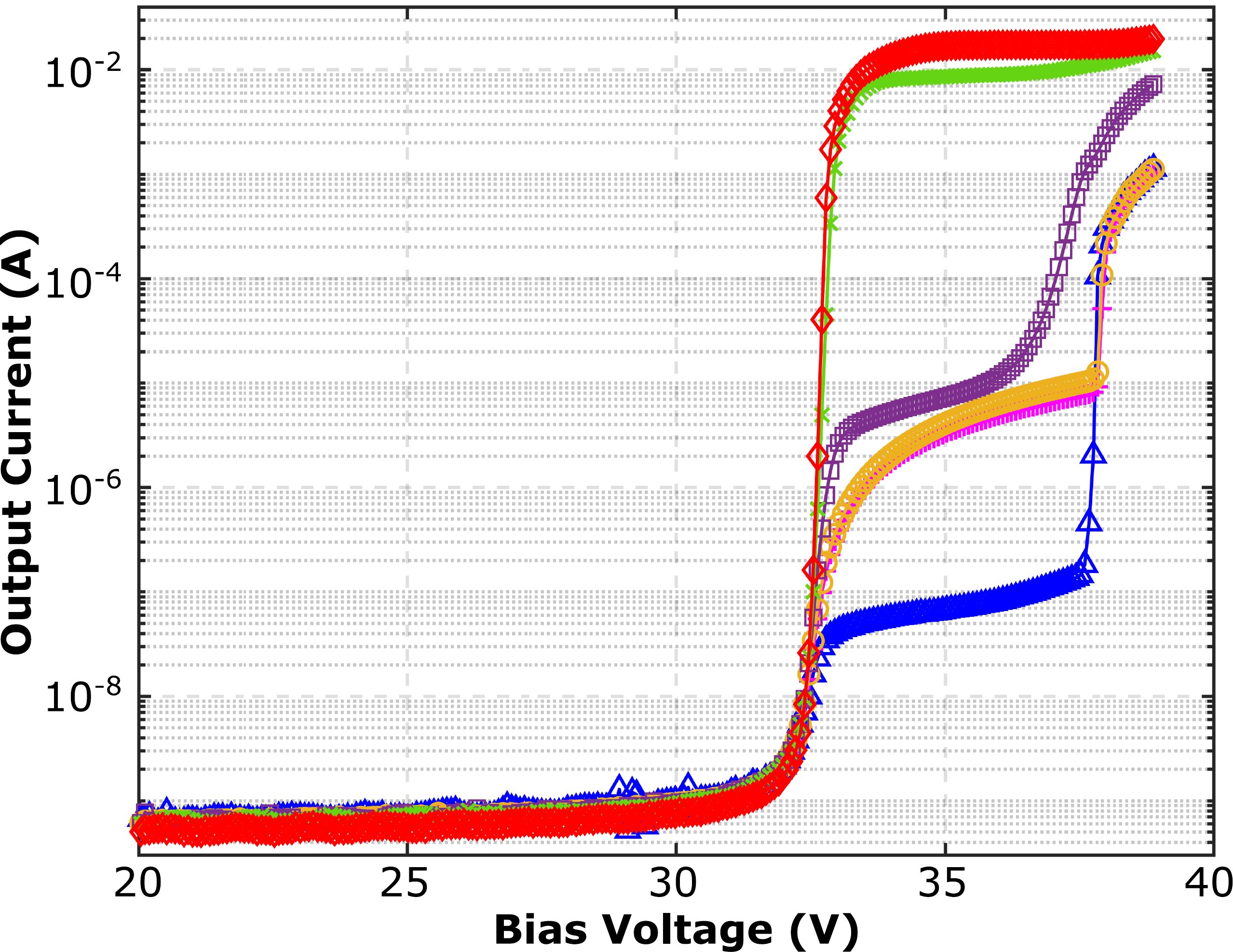}
\caption{I-V characteristics with different $V_q$ at room temperature.}
\label{fig_IV_curve}
\end{figure}

As a first step in the characterization, we performed the I-V measurement of the SiPM. Fig. \ref{fig_IV_curve} shows the results of this test. The measurement was performed using a source measurement unit (SMU), forcing a voltage between the cathode terminal of the device and the SiPM output (source of the quenching transistors), and measuring the output current. The applied voltage was swept from 0 V to 38 V in order to cover all the operating points of the device. Moreover, various $V_q$ were used to check the influence of the quenching load on the behavior of the device. As we can notice from Fig. \ref{fig_IV_curve}, decreasing the value of $V_q$ (i.e., increasing the value of the quenching resistance) the voltage required to see an abrupt increment of the current, caused by the avalanche breakdown of all the devices in the array, increases. This is due to the voltage drop across the resistance that limits the actual voltage applied across the SPAD junction. Indeed, when $V_q$ = 3.3 V, the $I-V$ shows a behavior analogous to the one observed in \cite{gramuglia} for the single SPAD, showing a breakdown voltage of about 32 V. 
\subsection{Charge Spectrum Performance}\label{section_charge_spectrum}
The measurement setup used for charge spectrum and timing jitter characterization is shown in Fig~\ref{fig_schematic}c. A 780-nm laser pulsed at 10 MHz repetition rate and with an optical pulse width of 40 ps FWHM (ALS PiL079XSM) illuminates the whole SiPM area by means of a diffuser, while the light intensity on the array is adjusted by using neutral density filters. The quenching transistor is controlled with a voltage of 1.5 V. The output signal from the SiPM is taken across a 50 $\Omega$ load resistor, and amplified by a commercial amplifier (Mini Circuits ZKL-2R5+). The amplified signal is detected by an oscilloscope (Teledyne LeCroy WaveMaster 813 Zi-B), where the acquisition is triggered by the rising edge of the laser trigger pulses.
\begin{figure}
\centering
\includegraphics[width=1 \columnwidth]{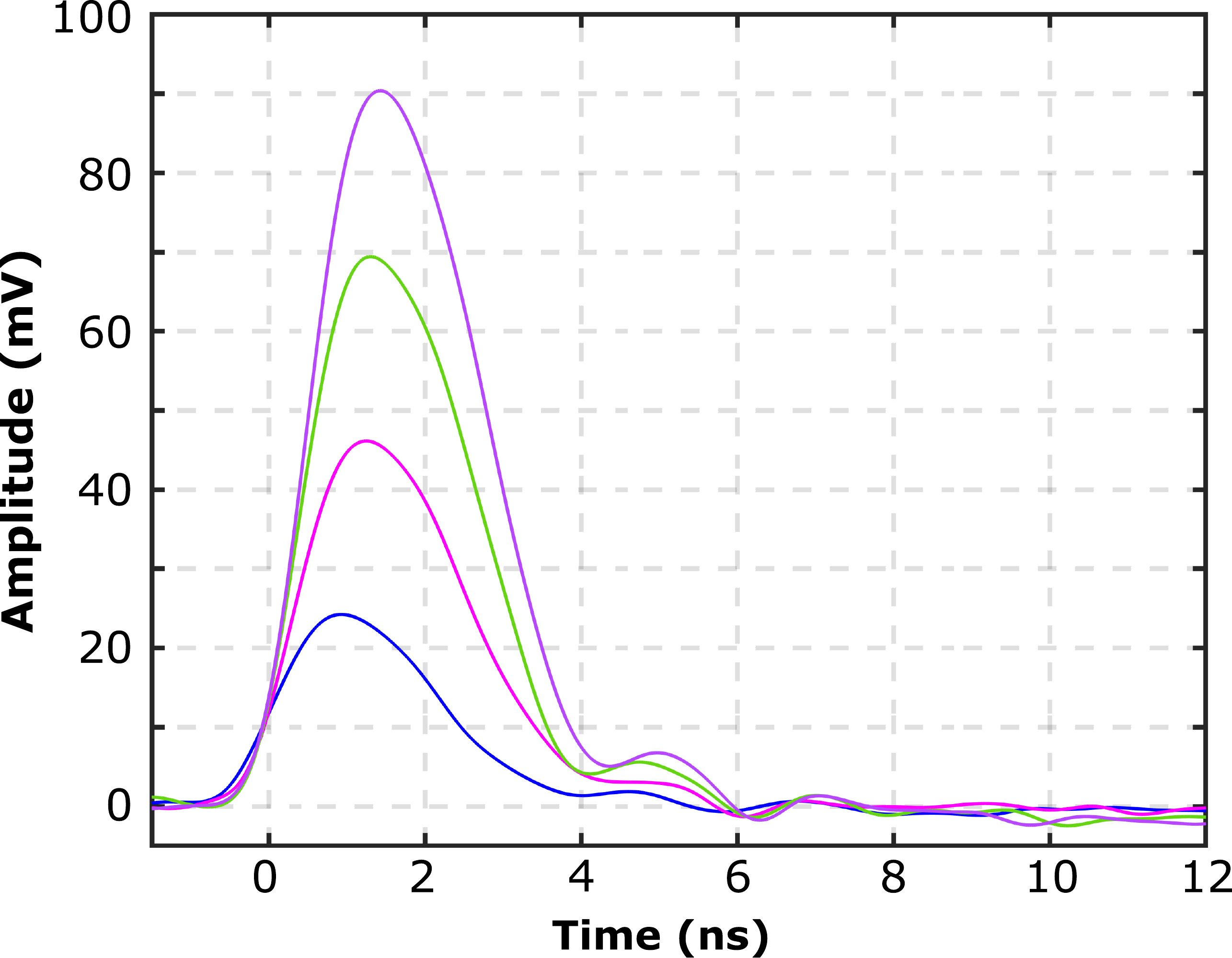}
\caption{Amplified output pulses corresponding to 1 pe, 2 pe, 3 pe and 4 pe events with 8 mV trigger threshold in oscilloscope.}
\label{fig_waveform}
\end{figure}

The amplified pulse shape of the SiPM output signal is shown in Fig.~\ref{fig_waveform}. The amplitude of the pulses is proportional to the number of simultaneously fired SPAD cells, which can be used to ascertain the number of detected photons. The pulse width of the signal is about 2.5 ns and the amplified signal from a single microcell is around 24 mV.

\begin{figure}
\centering
\includegraphics[width=1 \columnwidth]{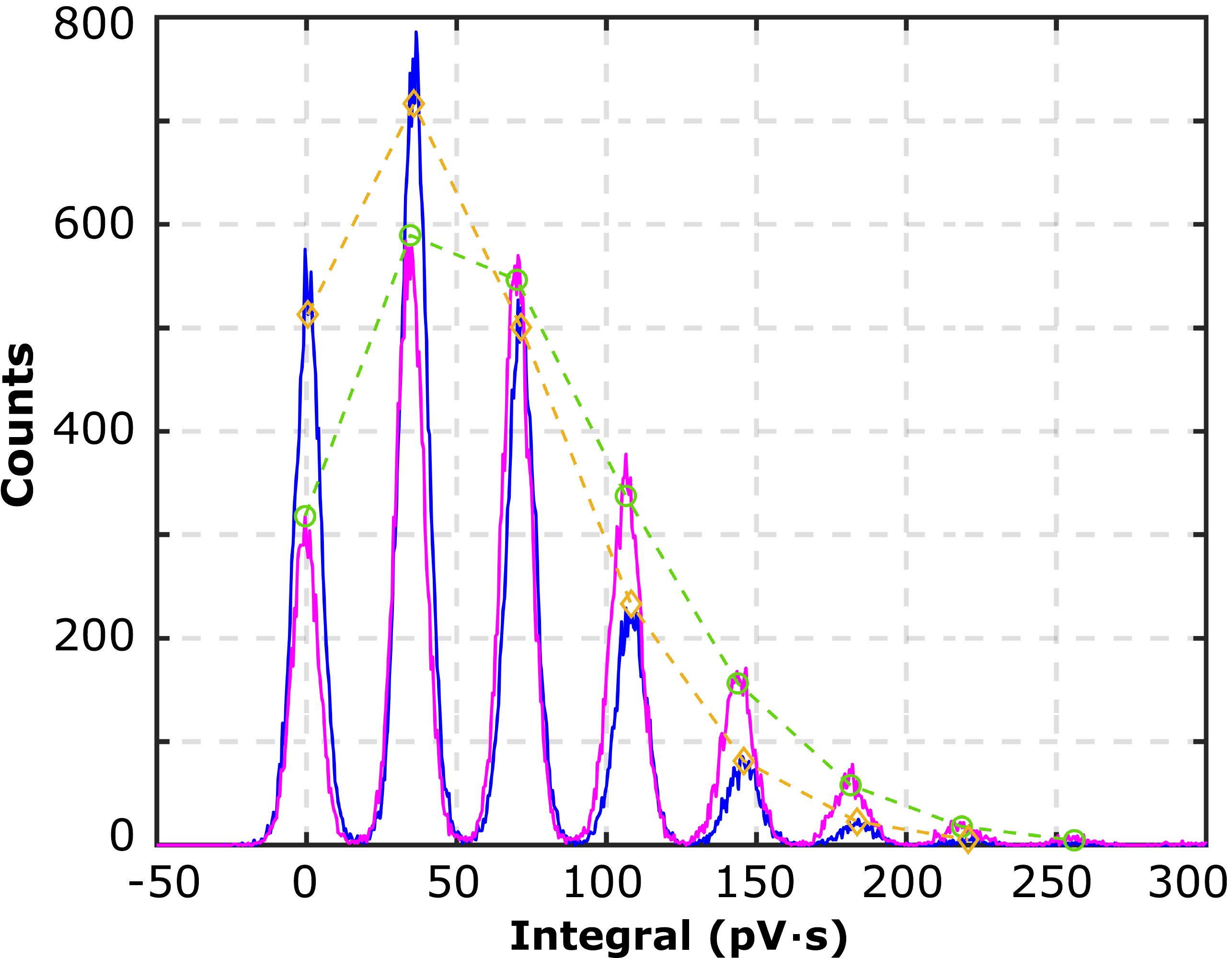}
\caption{Charge spectrum with low and high illumination conditions at 2V excess bias voltage with 50k events for each case.}
\label{fig_diagram_light}
\end{figure}

By extracting the area enclosed by the pulses generated from different fired cells, the charge spectrum can be acquired. In Fig.~\ref{fig_diagram_light} the envelope of the charge spectrum follows a Poisson distribution. Consequently, we can see that in Fig.~\ref{fig_diagram_light} the average number of events $\lambda$ under low illumination is 1.39 and 1.85 under high illumination after Poisson fitting. For this reason, the charge spectrum is an effective way to show the behavior of an a-SiPM as well as the capability of photon number resolving. In addition, the peak-to-valley ratio also reveals the correlated noise level in the device (i.e., afterpulsing and delayed crosstalk)~\cite{sanzaro20180}.

\begin{figure}
\centering
\includegraphics[width=1 \columnwidth]{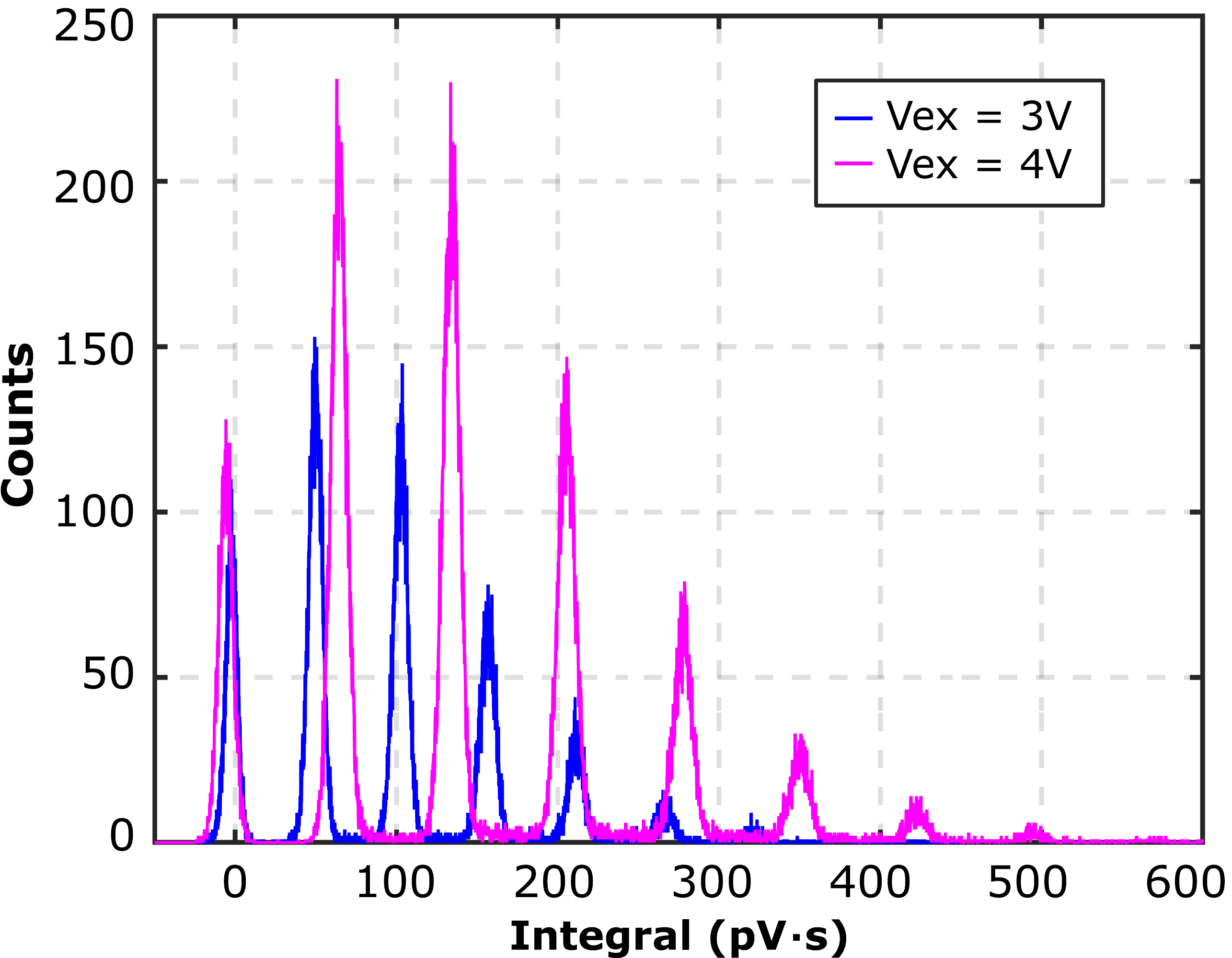}
\caption{Charge spectrum with two different excess bias voltages under similar light illumination conditions.}
\label{fig_diagram_vex}
\end{figure}

The gain of the a-SiPM is estimated by the charge spectrum. In fact, the difference between two neighboring peaks represent the gain as a function of applied bias voltages. As shown in Fig.~\ref{fig_diagram_vex}, the interval distances between two peaks increases with higher excess bias voltage. The gain is generally given by~\cite{acerbi2019understanding}
\begin{equation}\label{gain}
Gain = \frac{Q}{q} = \frac{V\cdot s}{R_L\cdot A},
\end{equation}
where $Q$ is the accumulated charge from avalanche, $q$ the elementary charge $q=1.602\times10^{-19}$ C, $R_L$ the load resistor, which is 50 $\Omega$ in our case and $A$ denotes the amplification factor of the amplifier. We calculated $V\cdot s$ as the average of the distances between two successive peaks in the charge spectrum. Fig.~\ref{gain} shows the measured estimated gain related to the applied bias voltages, exhibiting good linearity. The crossing point between the linear fitting line and the bias voltage is another approach to derive the breakdown voltage. This point is around 32 V, which confirms the aforementioned breakdown voltage estimation performed with the \emph{I-V} curve.
\begin{figure}
\centering
\includegraphics[width=1 \columnwidth]{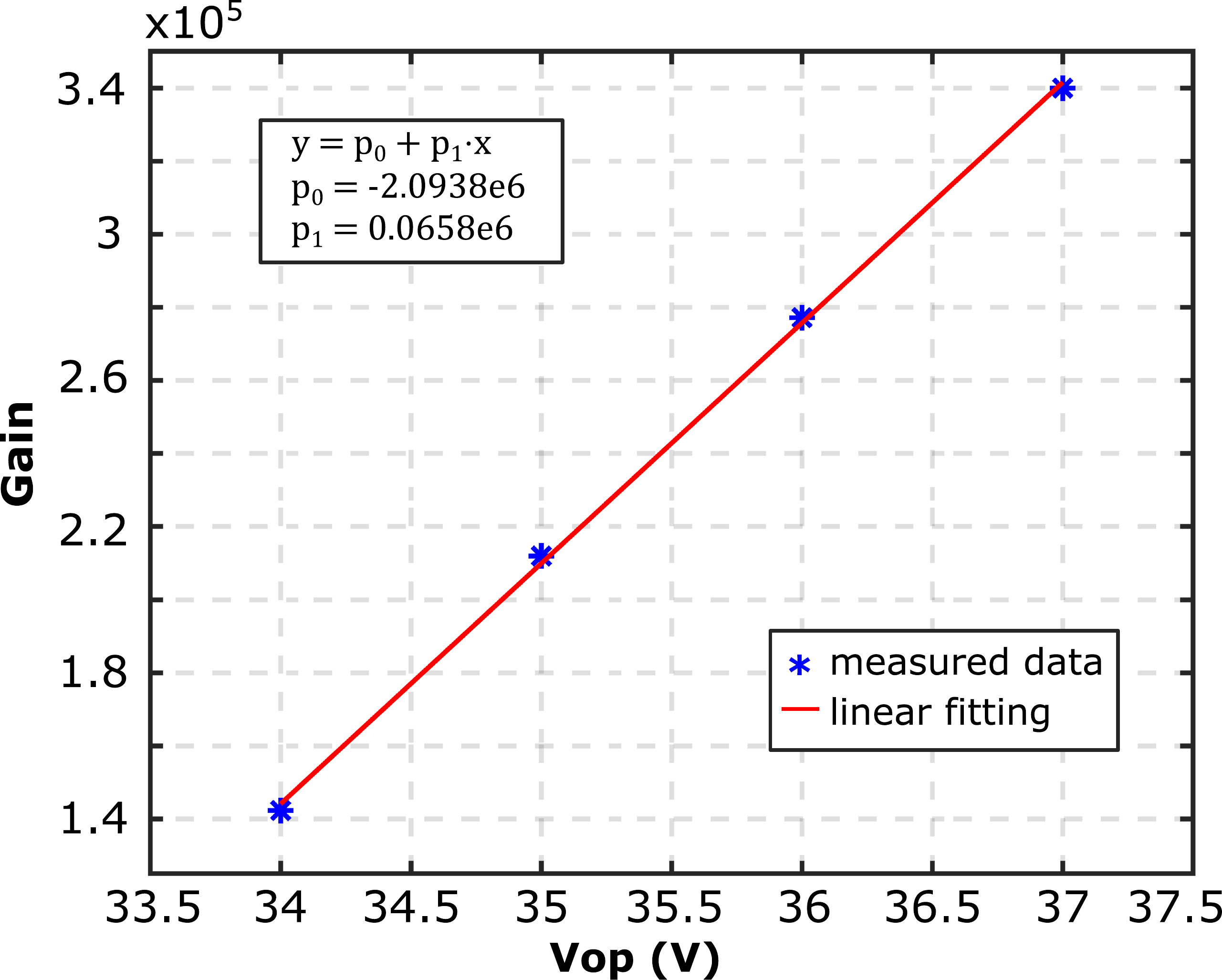}
\caption{Measured gain as a function of the applied bias voltages.}
\label{gain}
\end{figure}
\subsection{Jitter Performance}
\begin{figure}
\centering
\includegraphics[width=1 \columnwidth]{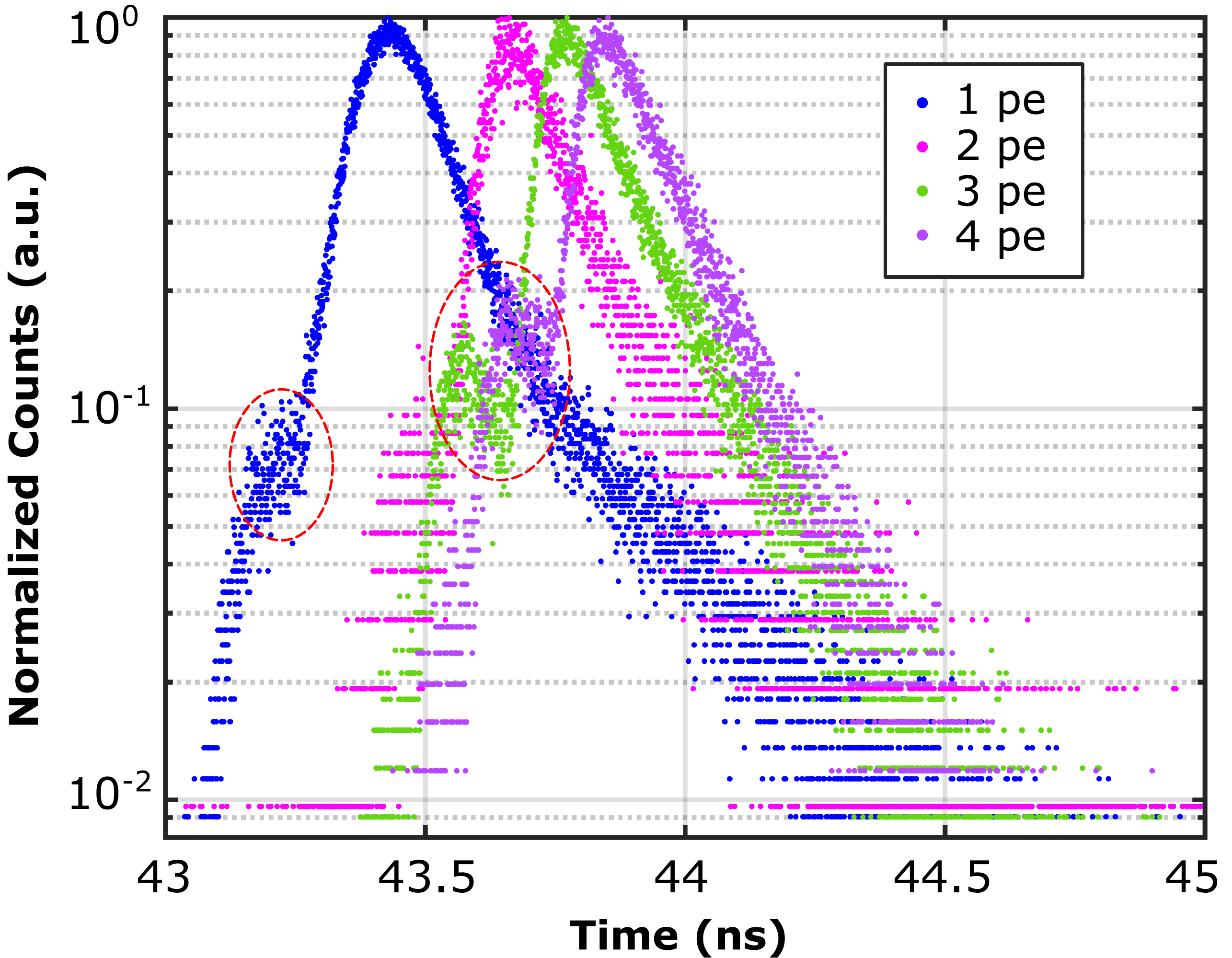}
\caption{SPTR and MPTR measurement results at excess bias voltage of 3.3V.}
\label{fig_hist}
\end{figure}

\begin{figure}
\centering
\includegraphics[width=1 \columnwidth]{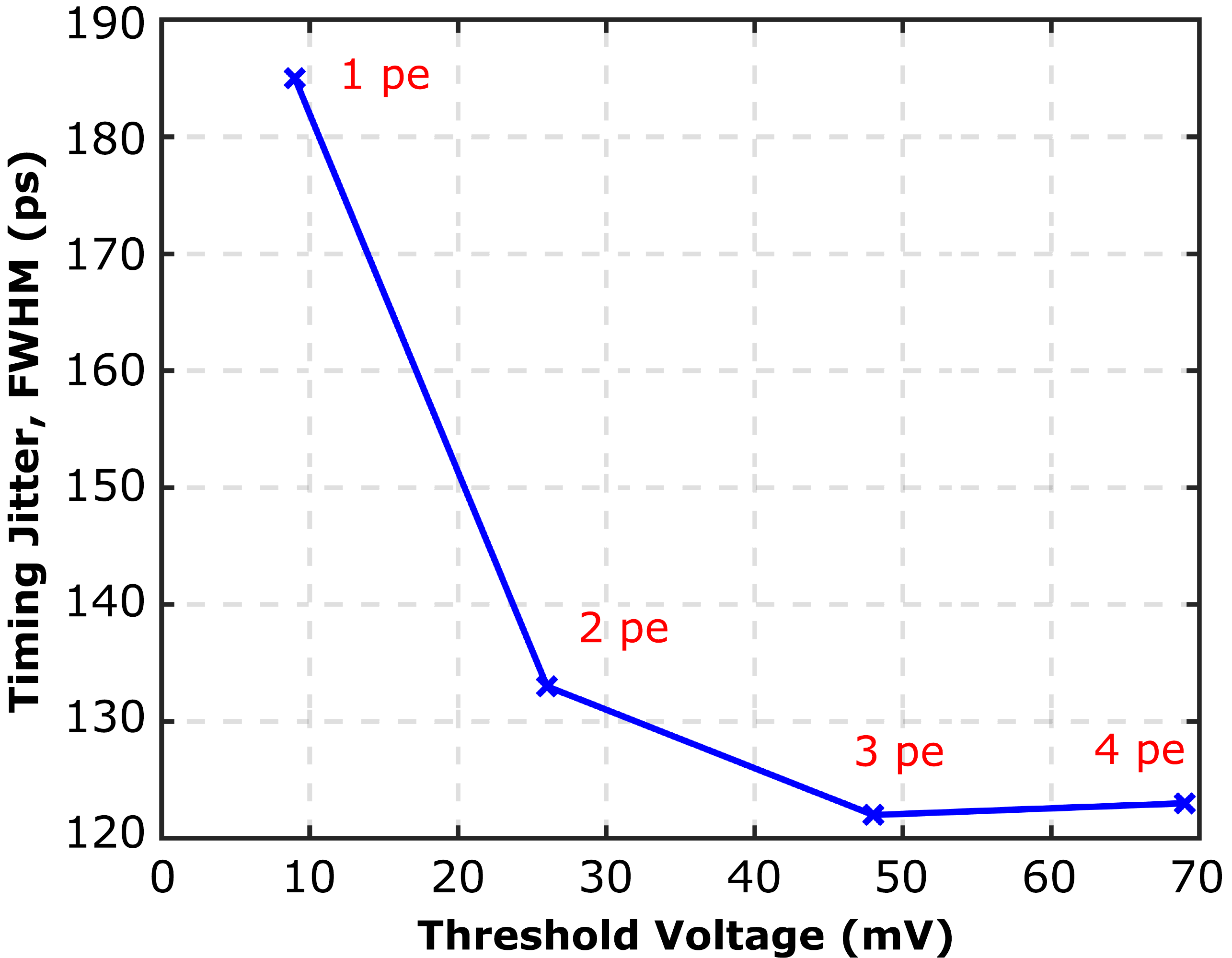}
\caption{The timing jitter at FWHM with excess bias voltage of 3.3V.}
\label{fig_jitter}
\end{figure}

The timing jitter performance is a critical parameter for LiDAR applications and it affects the precision of the measured distances. As the SiPM can resolve multiple photon events, the single-photon time resolution (SPTR) and multi-photon time resolution (MPTR) have been both measured in this work. We used the time-correlated single-photon counting (TCSPC) technique to accumulate the time differences between  laser trigger signal and  SiPM pulses. By adjusting the threshold voltage at the output of the amplifier, we can obtain SPTR and MPTR as shown in Fig.~\ref{fig_hist} with up to 4-photon events. Note that there are also some more photon events in each histogram as marked in red. The timing jitter values at full width at half maximum (FWHM) with different threshold voltages are plotted in Fig.~\ref{fig_jitter}. The SPTR is 185 ps and MPTR reaches a minimum with 3 photon events at 122 ps. As demonstrated by Fishburn and Seifert, and later by Gundacker this behavior can be explained by order-statistics and can be modeled using Fisher information~\cite{fishburn12,seifert2012lower,GUNDACKER20156}.

\section{LiDAR Demonstration}
\begin{figure}
\centering
\includegraphics[width=1 \columnwidth]{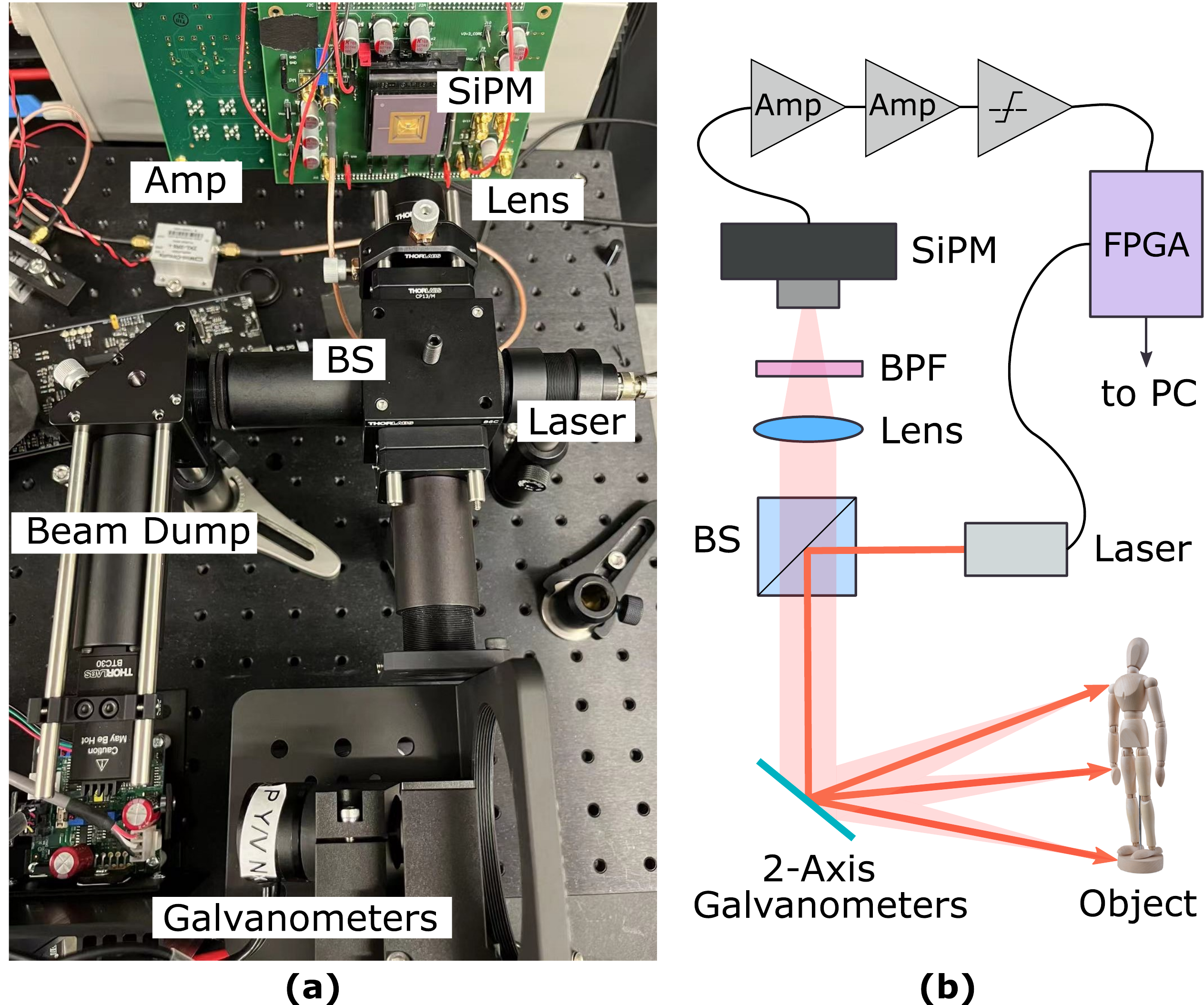}
\caption{The photograph and schematic of the co-axial LiDAR setup.}
\label{fig_lidar_setup}
\end{figure}

We integrate the SiPM into a co-axial LiDAR system as shown in Fig.\ref{fig_lidar_setup}. A collimated 780 nm laser (Picoquant VisIR) illuminates  the scene by way of a 2-axis galvanometer scanning mirror (Thorlabs GVS012). The reflected light is concentrated to the SiPM by a focal lens (f = 50 mm). A bandpass filter is placed in front of the SiPM to suppress the noise from ambient light. To further increase the pulse amplitude from the SiPM, one more commercial amplifier (Mini-Circuits ZFL-1000LN+) is added to the system. The amplified signal is discriminated by a comparator and then fed to a TCSPC module developed on FPGA (FELIX system) \cite{lusardi2021fpga,garzetti2021time,corna2021digital,lusardi2021cross}. The scanning and acquisition are synchronized in software.
\subsection{Dark Count Rate and Crosstalk Characteristics}
\begin{figure}
\centering
\includegraphics[width=1 \columnwidth]{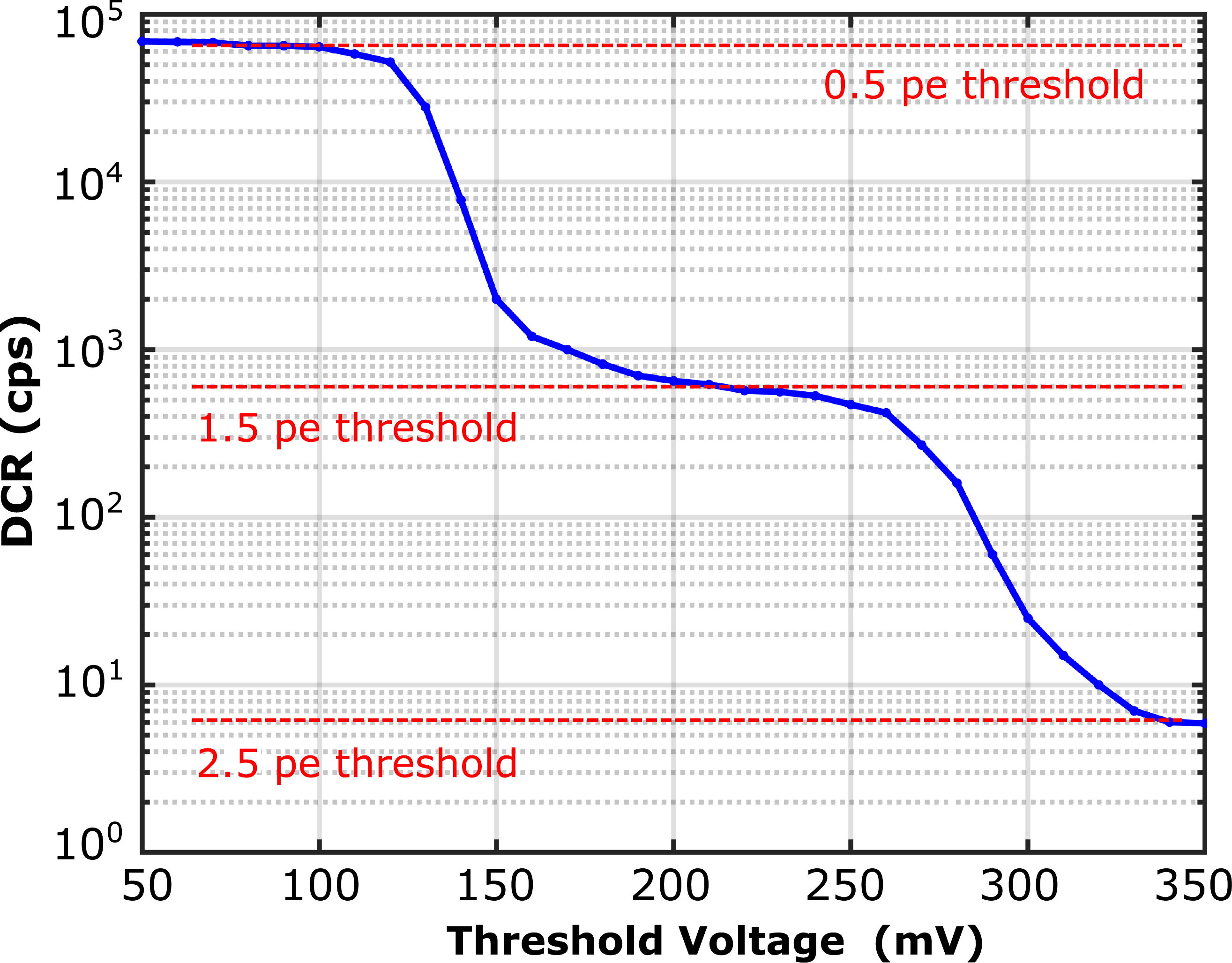}
\caption{DCR as a function of threshold voltage. The first reduction is achieved by ignoring hot SPADs (see definition in the text). All measurements were conducted at 3.3V of excess bias.}
\label{DCR}
\end{figure}

To characterize the dark count rate (DCR) and crosstalk probability, we put the device in the dark  and count the pulses using a counter implemented in FPGA by adjusting the threshold voltage of the comparator  from 50 mV to 350 mV. The excess bias voltage was fixed to 3.3 V, which is also the condition we used for range and LiDAR measurements. As shown in Fig.~\ref{DCR}, the DCR behaves as a step function according to the threshold voltages. The DCR in the plot is consistent with the DCR measured on individual SPADs of the same kind~\cite{gramuglia}  under comparable environmental conditions, whereas on average the DCR of each SPAD would be 390 cps at room temperature and 3.3 V of excess bias. This value also accounts for hot SPADs, defined as devices with at least 2 orders of magnitude higher DCR than the median value. By increasing the threshold beyond 1 pe, the DCR is reduced 2 orders of magnitude, indicating that the hot SPADs are ignored. We believe that a masking mechanism could enable us to permanently suppress these SPADs thereby reducing DCR accordingly. This will be the goal of our follow-on design~\cite{Zhang2019}.
By comparing the DCR above 1.5 pe threshold with the total measured DCR (0.5 pe threshold)~\cite{eckert2010}, the crosstalk probability can be calculated to less than 1\%. 
\subsection{Range Measurement}\label{section_range_measurement}
We started with a single-point range measurement in order to characterize the precision and accuracy of the LiDAR system. A movable object covered with white paper under indoor light condition (200 lux) is illuminated by the collimated laser. The laser powered with 1 mW average power runs at a repetition frequency of 5 MHz, corresponding to an unambiguous detected range of 30 m. The measurement was conducted from 0 to 25 m with a step of 0.5 m. A commercial rangefinder was used to measure the ground truth. In this measurement, the time-to-digital converter (TDC) in the FPGA was configured with the resolution of 37.5 ps. To obtain better ambient light suppression, we set a 250 mV threshold voltage to only count multi-photon events. We acquired a large set of events and organized them in histograms, with an exposure time of around 5 seconds.

\begin{figure}
\centering
\includegraphics[width=1 \columnwidth]{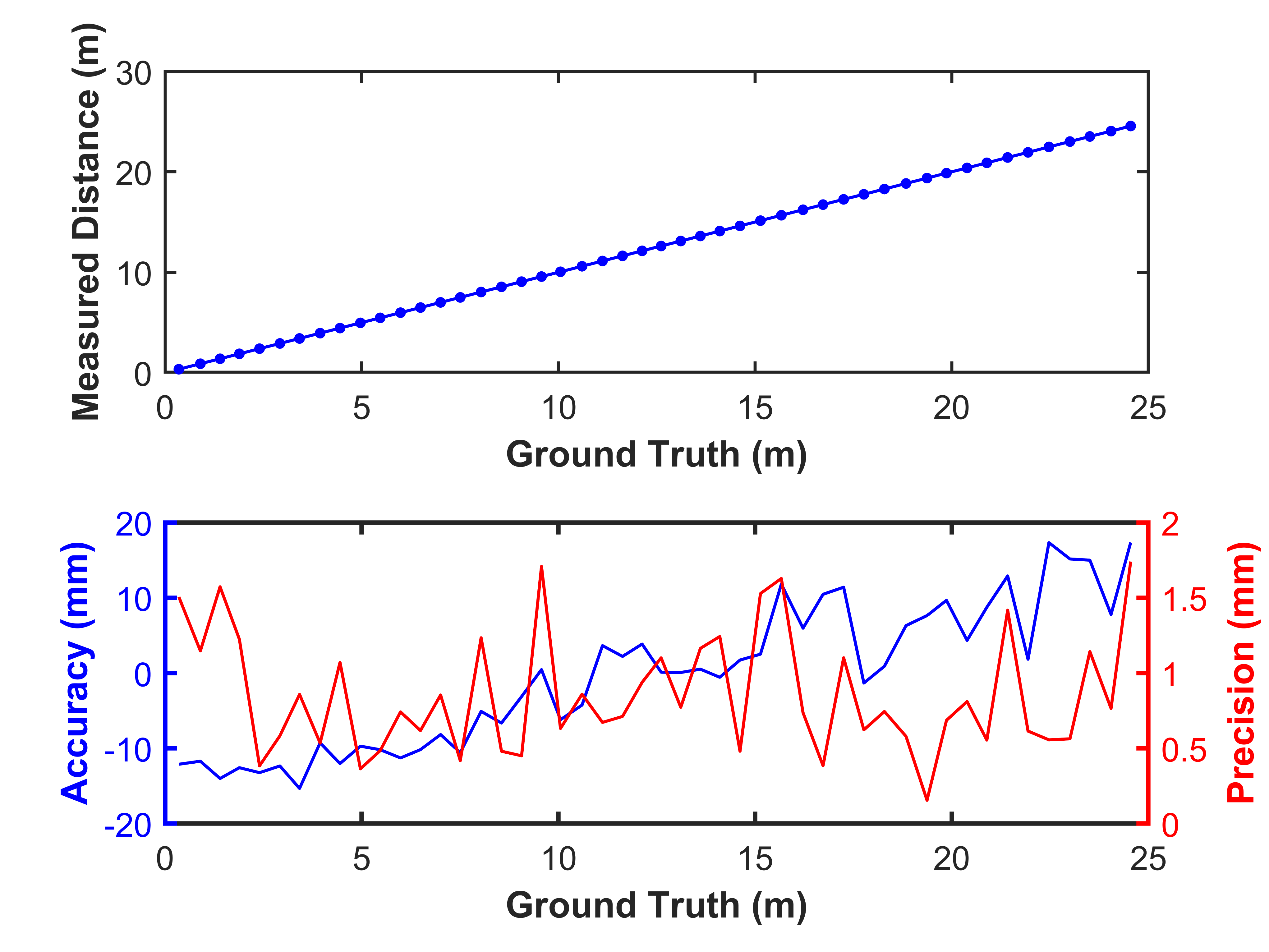}
\caption{Range measurement and accuracy and precision. The threshold was set so as to account for multi-photon events only.}
\label{fig_range}
\end{figure}

For each distance, 10 measurements were implemented, processed and then averaged to find the estimated distance value. The histogram processing is based on the center-of-mass (CoM) method. Here, we define the accuracy as the difference between the averaged distance value from 10 measurements to the ground truth, while the precision is the standard deviation of the measurement~\cite{zhang2021240}. The results of the measurements are shown in Fig. \ref{fig_range}. The LiDAR system achieves a precision better than 1.8 mm and an accuracy better than 1.85 cm over the whole range, in line with expectations \cite{Zhang2019, Ximenes2018}.
\subsection{Co-axial Scanning LiDAR}
\begin{figure*}[h]
\centering
\includegraphics[width=0.9 \textwidth]{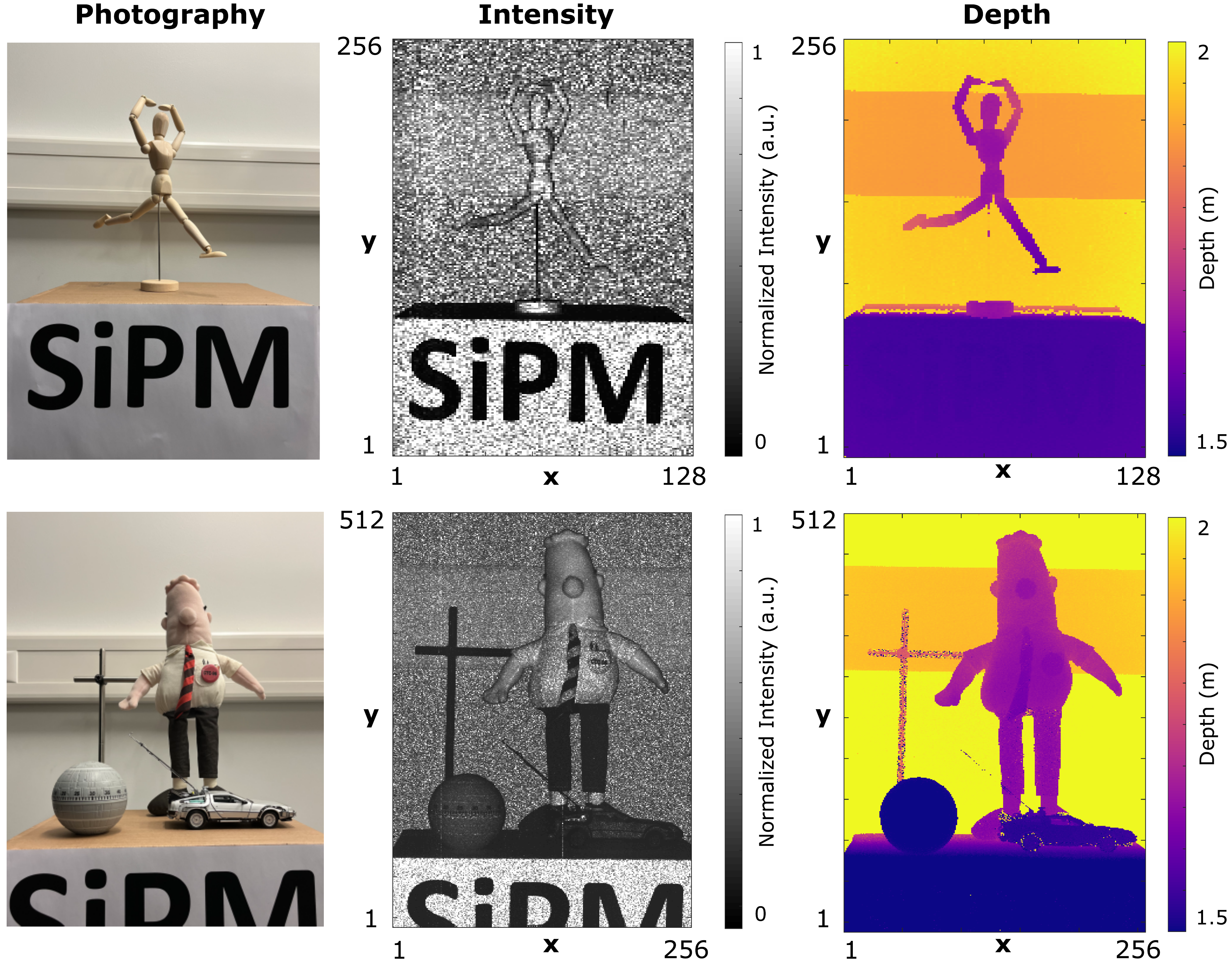}
\caption{Intensity and depth measurement results.}
\label{3D_image}
\end{figure*}
As shown in the photographs of Fig. \ref{3D_image}, we created two scenes with various albedos and materials for scanning LiDAR measurement with an ambient light of 100 lux. The objects were placed about 2 m away from the LiDAR system. The laser ran at the repetition frequency of 20 MHz with an average power of 2 mW. We scanned 256$\times$128 and 512$\times$256 points for the two scenes respectively. The configured threshold voltage for the comparator was set to 250 mV with an exposure time of 10 ms for each point in both scenes. Note that the minimum exposure time configuration is limited by the current software. While the frame rate could be further improved by reducing the acquisition time.

The results of the measurement with both intensity and depths are shown in Fig. \ref{3D_image}. In the intensity images, the objects can be well distinguished. However, very few photons were acquired from the specular metal pillars, where the injected light is reflected to some particular directions out of the field of view instead of scattering to all directions. The depth information is also well detected by the LiDAR system. For example, we can clearly see the depth differences of the mannequin at different parts. There are also some complex structures in the scenes, such as the wheels and the power pickup for the flux capacitor in the DeLorean model car from the movie `Back to the Future'. Here, multiple reflections might happen, but they are interpreted correctly in the LiDAR. In some cases, the LiDAR system could not capture enough returning photons, thus resulting in a distance estimation dominated by background noise. Algorithms like pixel-wise smoothing could be used to improve the performance for such details.


\section{Conclusion}
In this work, we have reported the design, experimental characterization and LiDAR application of the first analog SiPM in a 55 nm BCD process. The SiPM can be extended to a larger arrays by integrating more SPAD microcells. Furthermore, the results will enable the development of a cost-effective LiDAR system-on-chip (SoC) by integrating front-end circuits, TCSPC module and other analog or digital blocks in the advanced technology. The current results demonstrate the suitability of the approach based on this a-SiPM.

\section*{Acknowledgment}
The authors would also like to thank GlobalFoundries for
access to and assistance with the technology. A special thanks give to the group of the professor Angelo Geraci (Politecnico di Milano) and TEDIEL S.r.l. (spin-off of Politecnico di Milano) that have provided the FELIX system used for the timestamp acquisition.

\ifCLASSOPTIONcaptionsoff
  \newpage
\fi



%
\bibliographystyle{IEEEtran}
\bibliography{references}

\end{document}